\documentclass[%
 reprint,
superscriptaddress,
nofootinbib,
 amsmath,amssymb,
 aps,
]{revtex4-2}

\usepackage{graphicx}
\usepackage{dcolumn}
\usepackage{bm}
\usepackage[dvipsnames]{xcolor}
\usepackage[colorlinks=true, pdfstartview=FitV, linkcolor=RedOrange, citecolor=ForestGreen, urlcolor=blue]{hyperref}

\renewcommand{\d}{\mathrm{d}}
\newcommand{\D}{\mathrm{D}}

\begin{document}

\title{An effective macroscopic description of early-stage chemical equilibration in high-energy heavy-ion collisions}

\author{Anar Akbarov}
\email{anar.akbarov@wayne.edu}
\affiliation{Department of Physics and Astronomy, Wayne State University, Detroit, MI 48201, USA}

\author{Hendrik Roch}
\email{hendrik.h.roch@jyu.fi}
\affiliation{Department of Physics, University of Jyv\"askyl\"a, P.O. Box 35, 40014 University of Jyv\"askyl\"a, Finland}
\affiliation{Helsinki Institute of Physics, P.O. Box 64, 00014 University of Helsinki, Finland}

\author{Chun Shen}
\email{chunshen@wayne.edu}
\affiliation{Department of Physics and Astronomy, Wayne State University, Detroit, MI 48201, USA}

\date{\today}

\begin{abstract}
The early stage of high-energy heavy-ion collisions undergoes transient dynamics, evolving non-perturbatively from a gluon-rich to a chemically equilibrated quark-gluon plasma. 
We develop an effective hydrodynamic approach that casts the microscopic quark equilibration dynamics as the relaxation evolution of an effective macroscopic bulk viscous pressure. 
The evolution of this effective bulk viscous pressure captures quark chemical equilibration and the breaking of conformal symmetry, thereby providing a smooth onset of the lattice QCD equation of state after a few fm/$c$ in the hydrodynamic evolution. 
We investigate phenomenological effects of this macroscopic description on final-state observables in $\mathrm{Pb}+\mathrm{Pb}$, $\mathrm{O}+\mathrm{O}$, and $\mathrm{p}+\mathrm{Pb}$ collisions at the Large Hadron Collider.
\end{abstract}

\maketitle


\section{Introduction}

High-energy heavy-ion collisions produce a strongly coupled quark-gluon system, which behaves as a nearly inviscid fluid shortly after the initial impact~\cite{Song:2010mg, Heinz:2013th, Bernhard:2019bmu, ALICE:2022wpn}. 
It remains one of the central theoretical challenges to understand the early-time dynamics of the collisions and the onset of hydrodynamics, usually referred to as ``hydrodynamization''~\cite{Gale:2013da, Busza:2018rrf, Berges:2020fwq, Shen:2020mgh, Schenke:2021mxx, Noronha:2024dtq}. 
The initial state of heavy-ion collisions is characterized by highly occupied gluonic fields and strong momentum-space anisotropies~\cite{Gelis:2009wh, Schenke:2012wb, Berges:2020fwq}. 
The system undergoes rapid momentum isotropization and chemical equilibration during the pre-equilibrium stage~\cite{Chesler:2010bi, Heller:2012km, Kurkela:2015qoa, Kurkela:2018wud, Kurkela:2018xxd}.

The non-perturbative nature of the transition from an initially gluon-rich plasma to a chemically equilibrated Quark-Gluon Plasma (QGP) makes it particularly challenging to model~\cite{Keegan:2015avk}. 
The highly occupied gluon fields at the earliest stage of the collision are far from equilibrium and may not admit a well-defined quasiparticle description~\cite{Baier:2000sb}. 
While weak-coupling effective kinetic theories provide valuable insights into thermalization dynamics, most studies have assumed conformal symmetry, in which quarks and gluons are treated as effectively massless degrees of freedom~\cite{Liu:2015nwa, Kurkela:2018wud, Kurkela:2018vqr, Schenke:2020mbo, Du:2025hyk}.
In realistic strongly interacting systems, finite quark masses, the running of the Quantum Chromodynamics (QCD) coupling, and non-perturbative medium effects generate sizable deviations from conformality, especially near the QCD crossover region, where the system's trace anomaly is comparable to its thermal pressure~\cite{Borsanyi:2013bia, HotQCD:2014kol}. 
As a consequence, it remains difficult to quantitatively assess how breaking conformal symmetry modifies quark production rates, chemical equilibration processes, and subsequent hydrodynamic evolution. 
These uncertainties become especially important when one attempts to incorporate non-conformal dynamics and bulk viscous effects consistently into phenomenological simulations~\cite{daSilva:2022xwu}.

In many phenomenological studies of relativistic heavy-ion collisions, a positive initial bulk viscous pressure has often been introduced to account for deviations from conformality when transitioning from the pre-equilibrium phase to hydrodynamics~\cite{Bernhard:2019bmu, Schenke:2019pmk, Schenke:2020mbo, JETSCAPE:2020mzn, Nijs:2020roc}. The matching between the pre-equilibrium and hydrodynamic stages can itself generate sizable initial bulk viscous pressures, with the magnitude depending on how the distinct equations of state are matched~\cite{Borghini:2024kll}.
It was shown that this choice of initial bulk viscous pressure produced sizable changes in radial flow, particle spectra, and flow observables, particularly in small collision systems where the expansion rate is large, and the system spends a significant fraction of its lifetime far from equilibrium~\cite{Schenke:2019pmk, Chiu:2021muk, daSilva:2022xwu, Chiu:2025mau}. 
Beyond its phenomenological importance, the microscopic origin of bulk viscosity during chemical equilibration remains poorly understood, and a first-principles connection between chemical non-equilibrium and bulk viscous dynamics is still lacking.

Recently, Ref.~\cite{Gordeev:2025vog} proposed a dynamical framework in which the chemical equilibration process was modeled using a relaxation-type description, assuming the chemical equilibration time was independent of local thermodynamic variables. 
The authors quantified the impact of chemical equilibration dynamics by propagating out-of-chemical-equilibrium effects in the equation of state into the collision system's evolution and radiated electromagnetic (EM) observables. 
In practice, such an approach provides a tractable phenomenological implementation. 
However, it remains model-dependent how to connect different equations of state during the pre-equilibrium evolution from a gluon-rich system to chemically equilibrated quark-gluon matter~\cite{Vovchenko:2016ijt}. 
Different model realizations may obscure the thermodynamic origin of bulk viscous effects and the interplay between chemical non-equilibrium and conformal symmetry breaking in the pre-equilibrium stage.

The present work is motivated by the recently discovered duality between bulk viscous pressure and chemical potentials developed in Ref.~\cite{Gavassino:2023xkt}. 
This duality provides a novel macroscopic framework to study the system's chemical equilibration processes as effective bulk viscous corrections. 
The formalism was subsequently applied to quantify bulk viscous effects in neutron star mergers~\cite{Yang:2023ogo}, where chemical equilibration processes play an essential role in the dense QCD medium. 
Building on these developments, we explore how this duality can describe the chemical equilibration dynamics of the early-time QGP produced in relativistic heavy-ion collisions. 
Our approach provides a new description of chemical non-equilibrium and bulk viscous dynamics during the early stage of heavy-ion collisions, which improves the matching procedures between distinct equations of state.

The rest of the paper is organized as follows. 
Section~\ref{Sec:model} introduces the chemical rate equations for quark production and uses duality to map these microscopic processes to an effective bulk viscous pressure $\Pi_\mathrm{chem}$. 
Then, we carry out phenomenological studies with this framework and quantify the impact of $\Pi_\mathrm{chem}$ on final-state observables as a function of collision system size at Large Hadron Collider (LHC) energies in Sec.~\ref{Sec:Results}. 
Section~\ref{Sec:Conclusion} summarizes our results and concludes the paper.

\section{THEORETICAL FRAMEWORK}
\label{Sec:model}

\subsection{Chemical reactions for quark equilibration}

We model the chemical evolution of a gluon-rich medium using a phenomenological set of rate equations.
The dominant mechanism governing the change in quark-gluon composition is the $2 \leftrightarrow 2$ inelastic gluon fusion channel: $g + g \rightarrow q + \bar{q}$, along with its time-reversed counterpart, $q + \bar{q} \rightarrow g + g$.

The temporal evolution of the number densities for gluons ($n_g$), quarks ($n_q$), and antiquarks ($n_{\bar{q}}$) obeys a coupled system of non-linear rate equations:
\begin{align}
\frac{\d n_g}{\d t} &= -\sigma\left(\frac{n_g^2}{2} - n_q n_{\bar{q}}\right),\label{eq:dng_dt} \\
\frac{\d n_q}{\d t} &= \sigma\left(\frac{n_g^2}{2} - n_q n_{\bar{q}}\right), \\
\frac{\d n_{\bar{q}}}{\d t} &= \sigma\left(\frac{n_g^2}{2} - n_q n_{\bar{q}}\right).\label{eq:dnqbar_dt}
\end{align}
Here, $\sigma$ denotes the effective cross section for converting gluon pairs into quark--antiquark pairs.
These reaction equations ensure particle number conservation $2 \d n_g/\d t + \d n_q/\d t + \d n_{\bar{q}}/\d t = 0$ and baryon number conservation $\d n_q/\d t - \d n_{\bar{q}}/\d t = 0$. 
Detailed balance is achieved when $n_g^2 = 2 n_q n_{\bar q}$ and the net rate vanishes.

Within a dynamically expanding QGP created in relativistic heavy-ion collisions, we consider particle number currents $J_i^\mu = n_i u^\mu$ with $i\in\{g,q,\bar{q}\}$. 
Their evolution equations read,
\begin{align}
\D n_g + n_g \Theta &= -\sigma\left(\frac{n_g^2}{2} - n_q n_{\bar{q}}\right),  \label{eq:RateEq_ng}\\
\D n_q + n_q \Theta &= \sigma\left(\frac{n_g^2}{2} - n_q n_{\bar{q}}\right),  \label{eq:RateEq_nq}\\
\D n_{\bar{q}} + n_{\bar{q}} \Theta &= \sigma\left(\frac{n_g^2}{2} - n_q n_{\bar{q}}\right). \label{eq:RateEq_nqBar}
\end{align}
Here, the convective derivative $\D$ is defined as $\D \equiv u^\mu \partial_\mu$. 
The system's expansion rate is $\Theta \equiv \partial_\mu u^\mu$. 
The $n_i \Theta$ terms account for the kinematic dilution of local particle densities due to volume expansion.

To track the system's progress of chemical equilibration, we define the dimensionless composition variable:
\begin{equation}
Y_q \equiv \frac{2 n_q n_{\bar{q}}}{n_g^2}.
\end{equation}
As the system evolves from a gluon-rich environment to chemical equilibrium, this variable evolves from 0 to 1. 
At finite net baryon density, we introduce the variable $A_q \equiv n_{\bar{q}}/n_q$ to track the asymmetry between quark and antiquark number densities.
Based on the rate equations, Eqs.~\eqref{eq:RateEq_ng}-\eqref{eq:RateEq_nqBar}, we can derive the evolution equations for $Y_q$ and $A_q$,
\begin{align}
    \D Y_q &= R \sqrt{Y_q} (1 - Y_q) \left(\sqrt{\frac{1}{2}}\frac{1}{\sqrt{A_q}}(1 + A_q) + \sqrt{Y_q} \right),  \label{eq:RateEq_Yq} \\
    \D A_q &= R \sqrt{\frac{1}{2}} \frac{\sqrt{A_q}}{\sqrt{Y_q}}(1 - Y_q) (1 - A_q).   \label{eq:RateEq_Aq}
\end{align}
where $R \equiv n_g \sigma$ is the effective reaction rate. 
Equation~\eqref{eq:RateEq_Yq} has fixed points at $Y_q = 1$ (stable) at chemical equilibration and $Y_q = 0$ (unstable) in the purely gluonic limit. 
Equation~\eqref{eq:RateEq_Aq} has fixed points at $A_q = 1$ (stable) when the system has zero net quark density or $A_q = 0$ (unstable) when the system has no antiquarks. 
When the system reaches chemical equilibrium, $Y_q = 1$, the antiquark-to-quark ratio $A_q$ stops changing.

In the limit when $A_q = 0$, $n_{\bar{q}} = 0$ and $n_q$ is finite, and the ratio $\sqrt{Y_q/A_q} = \sqrt{2}n_q/n_g$ is positive, Eqs.~\eqref{eq:RateEq_Yq} and~\eqref{eq:RateEq_Aq} are well-behaved.

From now on, we will simplify our discussion to zero net baryon density, $n_q = n_{\bar{q}}$, and $A_q = 1$. 
In this case,
\begin{align}
    D Y_q = R \sqrt{Y_q} (1 - Y_q) \left(\sqrt{2} + \sqrt{Y_q} \right).\label{eq:RateEq_Yq_Aq1}
\end{align}

\subsection{Mapping $Y_q$ to an effective bulk viscous pressure}

In this section, we follow the duality relation proposed in Ref.~\cite{Gavassino:2023xkt} to map the chemical-composition variable $Y_q$ to a macroscopic effective bulk-viscous pressure $\Pi_\mathrm{chem}$. 
This bulk viscous pressure $\Pi_\mathrm{chem}$ captures the modifications of thermal pressure when the system is out of chemical equilibrium,
\begin{equation}
    \Pi_\mathrm{chem}(e, Y_q) \equiv P(e, Y_q) - P_\mathrm{eq}(e).
    \label{eq:bulkChemDef}
\end{equation}
Here, $e$ is the system's local energy density, and $P_\mathrm{eq}(e)$ denotes the system's equation of state at chemical equilibrium.
Equivalent to thermal pressure, we can express the effective bulk viscous pressure $\Pi_\mathrm{chem}$ in terms of the system's trace anomaly $I \equiv e - 3P$,
\begin{align}
    \Pi_\mathrm{chem}(e, Y_q) = \frac{1}{3}\left[I_\mathrm{eq}(e) - I(e, Y_q) \right].
\end{align}
This way, the evolution of $\Pi_\mathrm{chem}$ is directly related to the degree of broken conformal symmetry in the system.

To construct the equation of state in the undersaturated case ($Y_q < 1$), we introduce the following interpolation for the trace anomaly
\begin{align}
    I(e, Y_q) = (1 - \sqrt{Y_q}) I_0(e) + \sqrt{Y_q} I_\mathrm{eq}(e),
    \label{eq:I_interp}
\end{align}
where $I_0(e)$ is the trace anomaly for a thermalized pure-gluon system.\footnote{Here we choose a linear parameterization in terms of the quantity $\sqrt{Y_q}$ which ensures the derived bulk relaxation time $\tau_{\Pi_\mathrm{chem}}$ remains finite when $Y_q = 0$.}
In this case, the effective bulk viscous pressure becomes:
\begin{equation}
\Pi_\mathrm{chem} = \frac{1}{3} (1-\sqrt{Y_q}) [I_\mathrm{eq}(e) - I_0(e)].
\label{eq:bulkChem_I}
\end{equation}

\subsection{Equation of motion for the effective bulk viscous pressure $\Pi_\mathrm{chem}$}

Following a similar procedure as in Ref.~\cite{Yang:2023ogo}, we can derive the evolution equation for $\Pi_\mathrm{chem}$,
\begin{align}
    \D\Pi_\mathrm{chem} = \frac{\partial \Pi_\mathrm{chem}}{\partial e}\bigg\vert_{Y_q} \D e + \frac{\partial \Pi_\mathrm{chem}}{\partial Y_q} \bigg\vert_{e} \D Y_q.
\end{align}
Using Eqs.~\eqref{eq:RateEq_Yq_Aq1} and~\eqref{eq:bulkChem_I}, and the hydrodynamic equation of motion, $\D e = - (e + P) \Theta$, we obtain
\begin{align}
    \D\Pi_\mathrm{chem} &+ \frac{R}{2} (1 + \sqrt{Y_q}) (\sqrt{2} + \sqrt{Y_q}) \Pi_\mathrm{chem}
= \nonumber \\
& - \frac{1}{3} (1 - \sqrt{Y_q}) \Big(\tfrac{\d I_\mathrm{eq}}{\d e}-\tfrac{\d I_0}{\d e}\Big)\,(e+P)\,\Theta.
\label{eq:BulkChemEvoEq}
\end{align}
When the system is static ($\Theta = 0$), Eq.~\eqref{eq:BulkChemEvoEq} reduces to a simple relaxation equation in which the effective bulk viscous pressure relaxes to zero at a rate $R(1 + \sqrt{Y_q})(\sqrt{2} + \sqrt{Y_q})/2$.

We can cast Eq.~\eqref{eq:BulkChemEvoEq} into the Israel–Stewart~\cite{Israel:1979wp} form,
\begin{equation}
    \tau_{\Pi_\mathrm{chem}} \D\Pi_\mathrm{chem} + \Pi_\mathrm{chem} = -\zeta_{\Pi_\mathrm{chem}} \Theta
    \label{eq:BulkChemEvoEqIS}
\end{equation}
where $\tau_{\Pi_\mathrm{chem}}$ is the bulk relaxation time and $\zeta_{\Pi_\mathrm{chem}}$ is the bulk viscosity coefficient. 
Eq.~\eqref{eq:BulkChemEvoEq} yields expressions for $\tau_{\Pi_\mathrm{chem}}$ and $\zeta_{\Pi_\mathrm{chem}}$
\begin{align}
    \!\!\tau_{\Pi_\mathrm{chem}} &= \frac{2}{R(1 + \sqrt{Y_q})(\sqrt{2} + \sqrt{Y_q})}, \\
    \!\!\zeta_{\Pi_\mathrm{chem}} &= \tau_{\Pi_\mathrm{chem}} \frac{1}{3} (1 - \sqrt{Y_q}) \left( \frac{\d I_\mathrm{eq}}{\d e} - \frac{\d I_0}{\d e} \right) (e + P).
\end{align}
By substituting the chemical compisition variable $Y_q$ with $\Pi_\mathrm{chem}$, we obtain
\begin{align}
    \tau_{\Pi_\mathrm{chem}} &= \frac{2}{R \left(2 - \frac{3 \Pi_\mathrm{chem}}{I_\mathrm{eq}(e) - I_0(e)}\right)\left(\sqrt{2} + 1 - \frac{3 \Pi_\mathrm{chem}}{I_\mathrm{eq}(e) - I_0(e)}\right)}, \label{eq:tauPi_Pichem} \\
    \zeta_{\Pi_\mathrm{chem}} &= \frac{2}{R \left(2 - \frac{3 \Pi_\mathrm{chem}}{I_\mathrm{eq}(e) - I_0(e)}\right) \left(\sqrt{2} + 1 - \frac{3 \Pi_\mathrm{chem}}{I_\mathrm{eq}(e) - I_0(e)}\right)} \nonumber \\
    & \quad \times \frac{\Pi_\mathrm{chem}}{I_\mathrm{eq} (e) - I_0(e)} \left( \frac{\d I_\mathrm{eq}}{\d e} - \frac{\d I_0}{\d e} \right) (e + P). \label{eq:zetaPi_Pichem}
\end{align}
Eq.~\eqref{eq:BulkChemEvoEq}, together with the effective transport coefficients in Eqs.~\eqref{eq:tauPi_Pichem}-\eqref{eq:zetaPi_Pichem}, forms a closed set of equations with the equation of motion for ideal hydrodynamics.

We note that, unlike the conventional Israel–Stewart evolution for the bulk viscous pressure whose transport coefficients are functions of only thermodynamic variables, the effective transport coefficients for $\Pi_\mathrm{chem}$ here are functions of $\Pi_\mathrm{chem}$. 
Therefore, the evolution of $\Pi_\mathrm{chem}$ in Eq.~\eqref{eq:BulkChemEvoEq} is expected to be highly non-linear.

\section{Results}
\label{Sec:Results}

In practice, we will use an equation of state based on lattice QCD smoothly connected with hadron resonance gas at chemical equilibrium, $I_\mathrm{eq}(e) = I_\mathrm{lQCD}(e)$~\cite{Moreland:2015dvc}, and assume a conformal ideal-gas equation of state in the early-stage pure gluon phase, $I_0(e) = 0$. 
Alternatively, one can adopt the lattice QCD equation of state for a pure-gluon system for $I_0(e)$~\cite{Borsanyi:2012ve}. 
Here, we impose conformal symmetry to be consistent with most initial-state and pre-equilibrium models based on effective kinetic theory (EKT)~\cite{Kurkela:2018vqr, Kurkela:2018wud, Du:2025hyk}.
We assume the effective chemical reaction rate $R = C_R T$, where $C_R$ is a dimensionless phenomenological parameter.

\subsection{Effective transport coeffcieints}

With our setup, we can compute the normalized (dimensionless) effective bulk viscosity related to $\Pi_\mathrm{chem}$,
\begin{align}
    \!\!\!\frac{\zeta_{\Pi_\mathrm{chem}}T}{e + P} = \frac{2(1 - \sqrt{Y_q})}{3 C_R (1 + \sqrt{Y_q}) (\sqrt{2} + \sqrt{Y_q})}  \left( \frac{\d I_\mathrm{lQCD}}{\d e} \right).
    \label{eq:zetaOvers}
\end{align}
A positive effective bulk viscosity in Eq.~\eqref{eq:zetaOvers} requires $\d I_\mathrm{lQCD}/\d e > 0$ when the system is gluon-rich, $Y_q \le 1$. 
As $Y_q$ evolves from 0 to 1, the normalized effective bulk viscosity reduces to zero when the system reaches chemical equilibrium, $Y_q = 1$.

\begin{figure}[t!]
  \includegraphics[width=\linewidth]{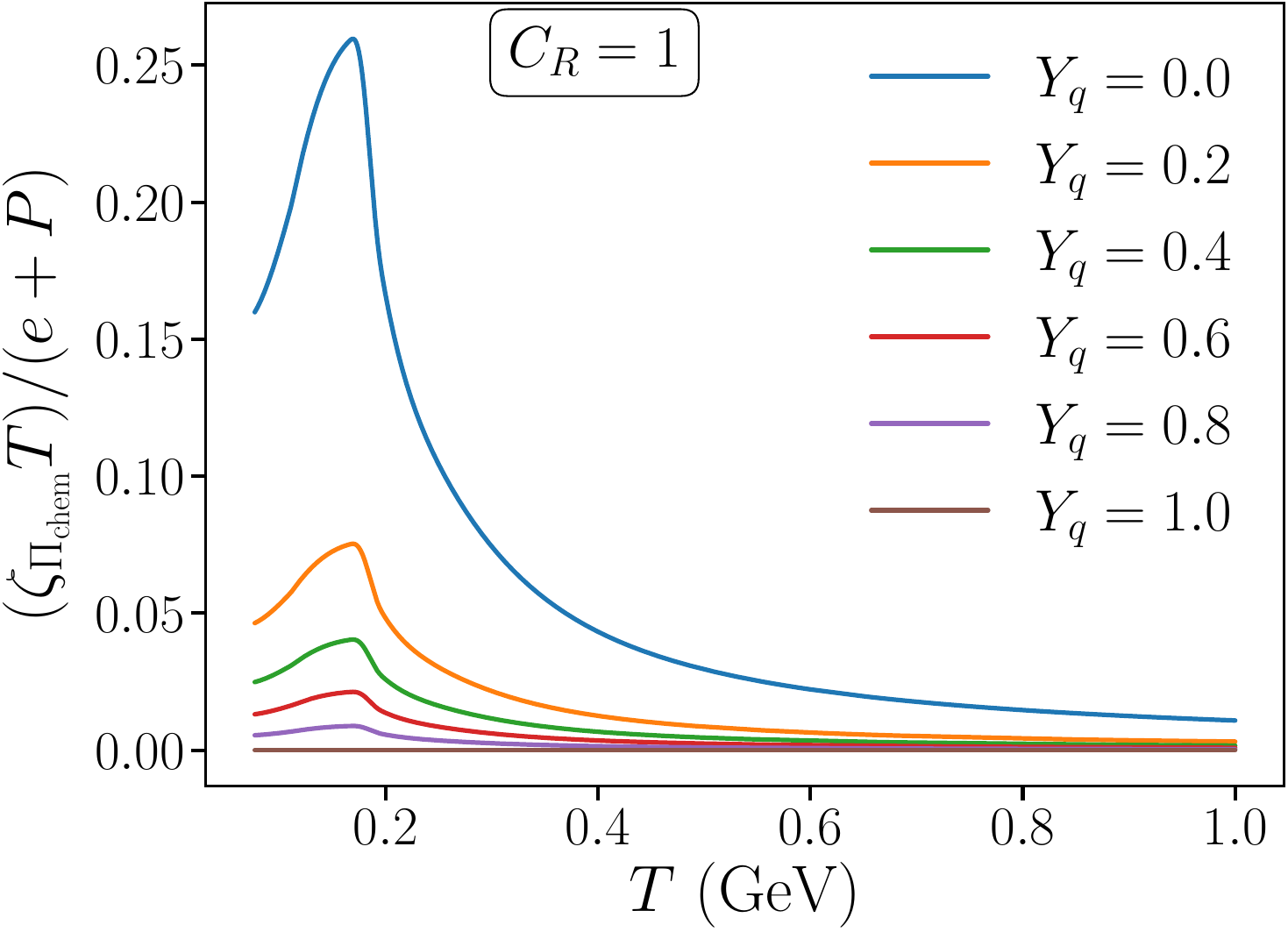}
  \caption{The normalized (dimensionless) effective bulk viscosity, arising from quark chemical equilibration with a reaction rate coefficient $C_R = 1$, as a function of temperature $T$ and the chemical composition variable $Y_q$.}
  \label{fig:zeta_over_s}
\end{figure}

Figure~\ref{fig:zeta_over_s} shows how the normalized effective bulk viscosity depends on temperature and the chemical composition variable $Y_q$ using the lattice QCD equation of state~\cite{Moreland:2015dvc}. 
According to Eq.~\eqref{eq:zetaOvers}, its temperature dependence is determined by the derivatives of the lattice equation of state, $\d I_\mathrm{lQCD}/\d e$. 
The normalized effective bulk viscosity reaches its maximum near the phase crossover temperature. 
Its maximum reaches around 0.25 when $Y_q = 0$, comparable to the specific bulk viscosity values extracted from phenomenological studies~\cite{JETSCAPE:2020mzn, Jahan:2024wpj, Jahan:2025cbp, JETSCAPE:2026hdw}.
As the system approaches chemical equilibrium, this effective bulk viscosity reduces to zero.

The effective bulk viscous relaxation time in our setting is
\begin{align}
    \tau_{\Pi_\mathrm{chem}} = \frac{2}{C_R(1 + \sqrt{Y_q})(\sqrt{2} + \sqrt{Y_q})} \frac{1}{T}.
    \label{eq:tauPi_Yq}
\end{align}
It remains positive and finite for $Y_q \ge 0$. 
Equation~\eqref{eq:tauPi_Yq} shows that $\tau_{\Pi_\mathrm{chem}}$ decreases by a factor of $2+\sqrt{2}$ while $Y_q$ increases from 0 to 1.

\begin{figure}[t!]
  \includegraphics[width=\linewidth]{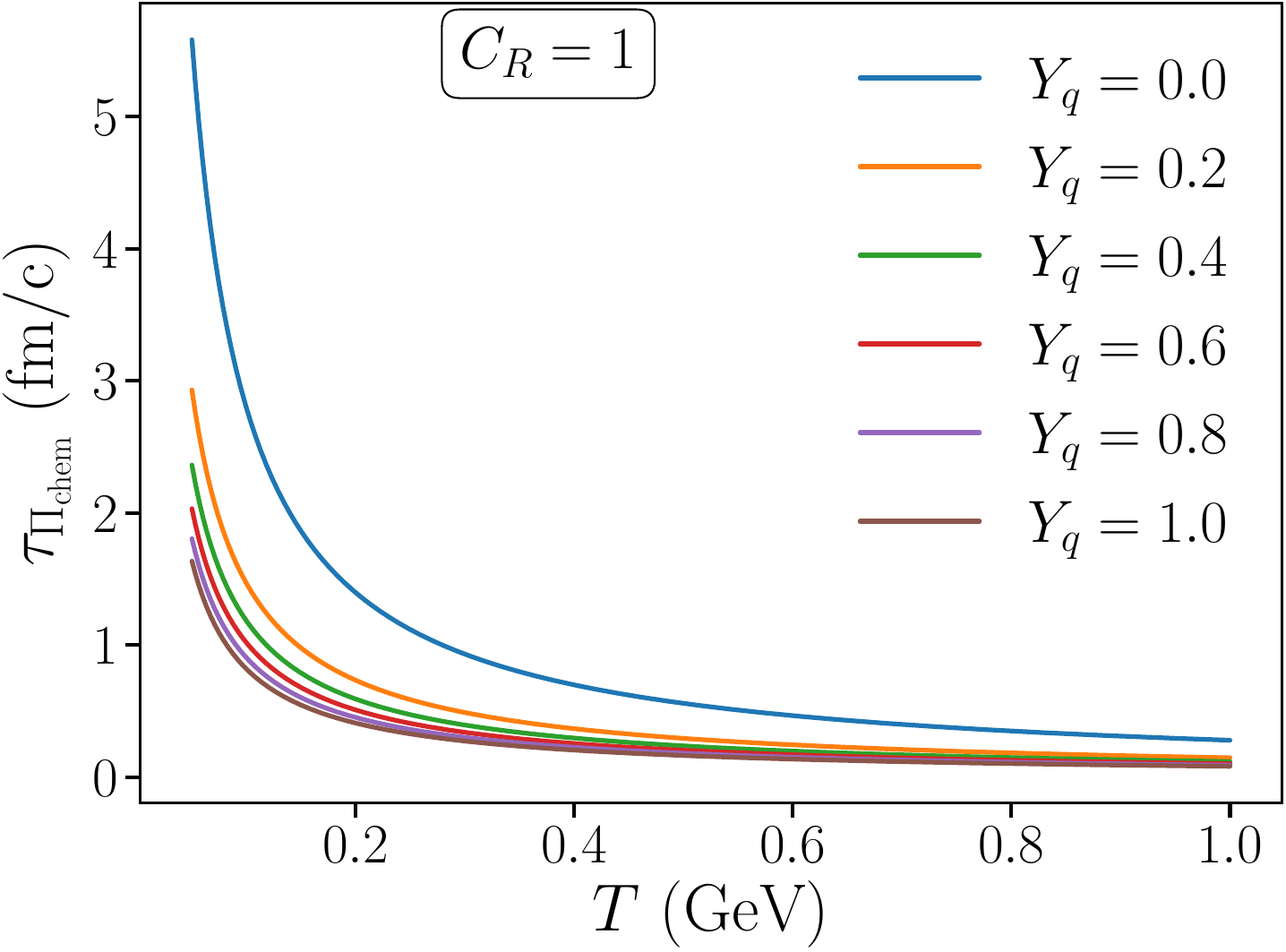}
  \caption{The relaxation time for the effective bulk viscosity, computed with a reaction rate coefficient $C_R = 1$, as a function of temperature $T$ and the chemical composition variable $Y_q$.} 
  \label{fig:tauPi}
\end{figure}

Figure~\ref{fig:tauPi} shows the effective bulk viscous relaxation time as a function of temperature and the chemical composition variable. 
It is independent of the equation of state. 
Within the temperature range probed by typical relativistic heavy-ion collisions, $\tau_{\Pi_\mathrm{chem}}$ remains on the order of 1 fm/$c$ with $C_R = 1$, comparable to the typical values of the viscous relaxation times for conventional shear and bulk viscosity used in phenomenological studies.

Overall, the effective transport coefficients are of similar magnitude to the conventional transport coefficients used in second-order viscous hydrodynamic simulations.

\subsection{Simulation setup and evolution of the effective bulk viscous pressure}

Now, we study the dynamical evolution of this new effective bulk viscous pressure $\Pi_\mathrm{chem}$ in event-by-event hydrodynamic simulations of relativistic heavy-ion collisions. 
We employ the IP-Glasma + JIMWLK + MUSIC + UrQMD framework~\cite{Mantysaari:2025tcg, Mantysaari:2025cos, Mantysaari:2025kls} to simulate collisions of different sizes at the LHC, assuming longitudinal boost invariance.
This framework first generates event-by-event gluon fields from the colliding nuclei at Bjorken $x = 0.01$ according to the IP-Sat model~\cite{Kowalski:2003hm}. 
These gluon fields are evolved under JIMWLK evolution to the appropriate small Bjorken-$x$ values at LHC energies~\cite{Schlichting:2014ipa, Mantysaari:2022sux} and collide with each other to produce the system's initial energy-momentum tensor $T^{\mu\nu}$ at $\tau = 0^+$.\footnote{We compute the Bjorken $x \equiv \langle p_{\mathrm{T}} \rangle/\sqrt{s_\mathrm{NN}}$ at mid-rapidity with an averaged $\langle p_{\mathrm{T}} \rangle = 0.65\;\mathrm{GeV}$~\cite{Mantysaari:2025tcg}.}
The system's energy-momentum tensor is then evolved using the classical Yang-Mills equations in the Glasma phase up to a proper time $\tau_0$, at which the gluon field strength is sufficiently weak to permit a transport description~\cite{Schenke:2012wb}. 
At $\tau_\mathrm{0}$, we match the system's energy-momentum tensor to the fields in relativistic viscous hydrodynamics in the Landau frame~\cite{Schenke:2020mbo}. 
The initial values of the effective bulk viscous pressure $\Pi_\mathrm{chem}$ are set as the trace-anomaly difference between the conformal and lattice QCD equations of state~\cite{HotQCD:2014kol}. 
The conventional bulk viscous pressure is initialized with $\Pi = 0$. 
After $\tau_0$, the system evolves with second-order viscous hydrodynamics~\cite{Paquet:2015lta} with the Denicol-Niemi-Moln\'{a}r-Rischke (DNMR) equations of motion~\cite{Denicol:2012cn},
\begin{align}
    \partial_\mu T^{\mu\nu} &= 0, \label{eq:ideal} \\
	\tau_{\Pi} \D \Pi + \Pi
	& = - \zeta \theta
	- \delta_{\Pi \Pi} \Pi \, \theta
	+ \lambda_{\Pi \pi} \pi^{\mu\nu} \sigma_{\mu\nu} \label{eq:bulk} \\ 
	\tau_{\pi} \D \pi^{\langle \mu\nu \rangle}
	+ \pi^{\mu\nu}
	& = 2\eta \sigma^{\mu\nu}
	- \delta_{\pi \pi} \pi^{\mu\nu} \theta
	+ \varphi_7 \pi_{\alpha}^{\langle \mu} \pi^{\nu\rangle \alpha} \nonumber \\
	& \quad - \tau_{\pi \pi} \pi_{\alpha}^{\langle \mu} \sigma^{\nu\rangle \alpha}
	+ \lambda_{\pi \Pi} \Pi\, \sigma^{\mu\nu}\,.
	\label{eq:shear}
\end{align}
We use the following temperature-dependent shear and bulk viscosities, which were calibrated to final-state observables in $\mathrm{Pb}+\mathrm{Pb}$ collisions at $\sqrt{s_\mathrm{NN}} = 5.02\;\mathrm{TeV}$ in Sec.~\ref{Sec:ResultsC}.
\begin{align}
    \frac{\eta}{s}(T) &= \left\{ \begin{array}{cc} \eta_0 & T > T_\eta \\
    \eta_0 + b (T - T_\eta) & T \le T_\eta \end{array} \right. ,\\
    \frac{\zeta}{s}(T) &= \left\{ \begin{array}{cc} \zeta_0 \exp\left[- \frac{(T - T_\zeta)^2}{2 \sigma_{\zeta+}^2} \right] & T > T_\zeta \\
    \zeta_0 \exp\left[- \frac{(T - T_\zeta)^2}{2 \sigma_{\zeta-}^2} \right] & T \le T_\zeta \end{array} \right. .
\end{align}
Here, the shear viscosity parameterization takes the parameters $\eta_0 = 0.12$, $T_\eta = 0.18\;\mathrm{GeV}$, and $b = -4\;\mathrm{GeV}^{-1}$.
For bulk viscosity, $\zeta_0 = 0.123$, $T_\zeta = 0.18\;\mathrm{GeV}$, $\sigma_{\zeta+} = 0.08\;\mathrm{GeV}$, and $\sigma_{\zeta-} = 0.03\;\mathrm{GeV}$.
The second-order transport coefficients take their standard values following Ref.~\cite{Denicol:2014vaa}.
When the system's local energy density drops below $e_\mathrm{sw} = 0.15\;\mathrm{GeV}/\mathrm{fm}^3$, individual fluid cells are converted into hadrons through the Cooper-Frye procedure~\cite{Huovinen:2012is, Shen:2014vra}. 
The thermally emitted hadrons are then fed to UrQMD~\cite{Bass:1998ca, Bleicher:1999xi} for hadronic scatterings and resonance decays.

\begin{figure}[h!]
  \includegraphics[width=\linewidth]{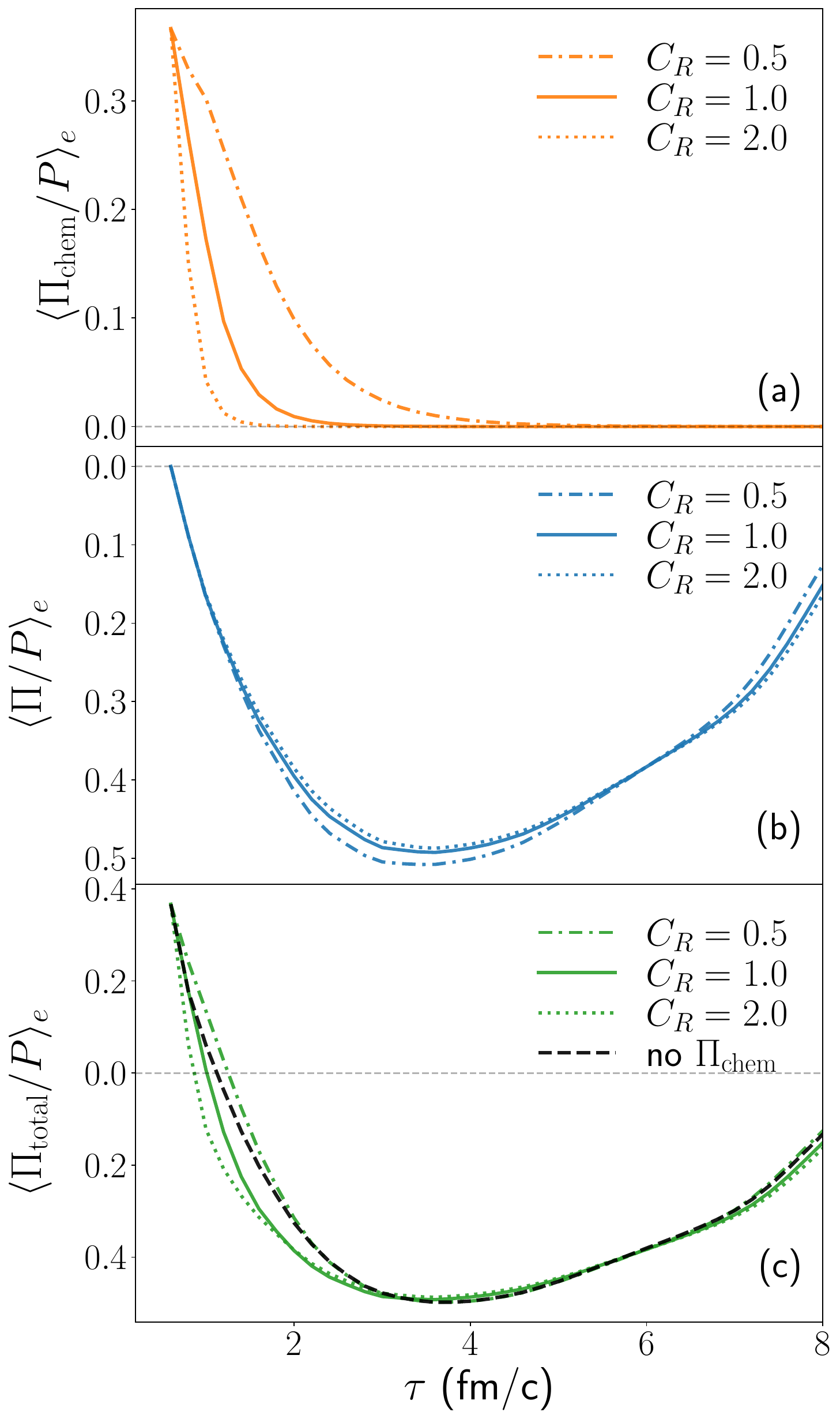}
  \caption{The time evolution of the effective bulk viscous pressure $\Pi_\mathrm{chem}$ (a), the conventional bulk viscous pressure $\Pi$ (b), and the total bulk viscous pressure $\Pi_\mathrm{total} = \Pi + \Pi_\mathrm{chem}$ (c), normalized by the thermal pressure $P$ with different values of $C_R$ for the reaction rates. The average is taken over the transverse plane, weighted by energy density (see text for definition).}
  \label{fig:bulk_chem}
\end{figure}
Figure~\ref{fig:bulk_chem} presents the time evolution of the effective bulk pressure $\Pi_\mathrm{chem}$, the conventional bulk viscous pressure $\Pi$, and the total bulk viscous pressure $\Pi_\mathrm{total}$, all normalized by the thermal pressure. 
The average is defined with local energy density $e(\mathbf{x})$ in the transverse plane as a weight, $\langle \mathcal{O} \rangle_e \equiv \int \d^2\mathbf{x}\,e(\mathbf{x}) \mathcal{O}(\mathbf{x}) /\int \d^2\mathbf{x}\,e(\mathbf{x})$. 
Because the effective bulk pressure $\Pi_\mathrm{chem}$ characterizes the pressure difference between the two equations of state as in Eq.~\eqref{eq:bulkChemDef}, $\langle \Pi_\mathrm{chem}/P \rangle_e$ starts positive and decreases exponentially in time as the system evolves towards chemical equilibrium. 
Its decay rate is controlled by the underlying chemical reaction rate $C_R$. Because $\Pi_\mathrm{chem}$ is positive, it increases the system's total pressure and enhances the flow development during the early stage of the hydrodynamic evolution.
On the other hand, the conventional bulk viscous pressure $\Pi$ acts as a resistance to local expansion. 
Its values are negative, driven by $-\zeta \theta$, thereby reducing the total pressure. 
As the system is accelerated by the local pressure gradients, the magnitude of the bulk viscous pressure $\Pi$ grows and reaches its maximum around $\tau = 3\;\mathrm{fm}/c$. 
The evolution of $\Pi$ shows a secondary dependence on the chemical reaction rate in Fig.~\ref{fig:bulk_chem}b. 
A small $C_R$ leads to a slower decay of $\Pi_\mathrm{chem}$, which generates more hydrodynamic flow, thereby increasing the magnitude of bulk viscous pressure $\Pi$.

Figure~\ref{fig:bulk_chem}c shows the evolution of the total bulk viscous pressure during the hydrodynamic expansion. 
Different chemical reaction rates result in sizable differences in $\Pi_\mathrm{total}$ during the first $3\;\mathrm{fm}/c$. 
Therefore, we expect the final-state observables in small collision systems with limited hydrodynamic lifetimes will show strong sensitivity to $C_R$. 
The black dashed line shows the bulk viscous pressure $\Pi$ in the standard setup, in which we treated pre-equilibrium chemical equilibration and collective expansion with one effective bulk viscosity. 
Its evolution lies between the cases with $C_R = 0.5$ and $C_R = 1$. 
Within our current formalism, we separate the bulk viscous effect in a chemically equilibrated QGP from the early-stage chemical equilibration process. 
Therefore, the phenomenologically constrained value is expected to be more precise compared to the one derived from the previous standard approach.

\subsection{Final State Observables}
\label{Sec:ResultsC}

\begin{figure}[t!]
  \centering
  \includegraphics[width=0.88\linewidth]{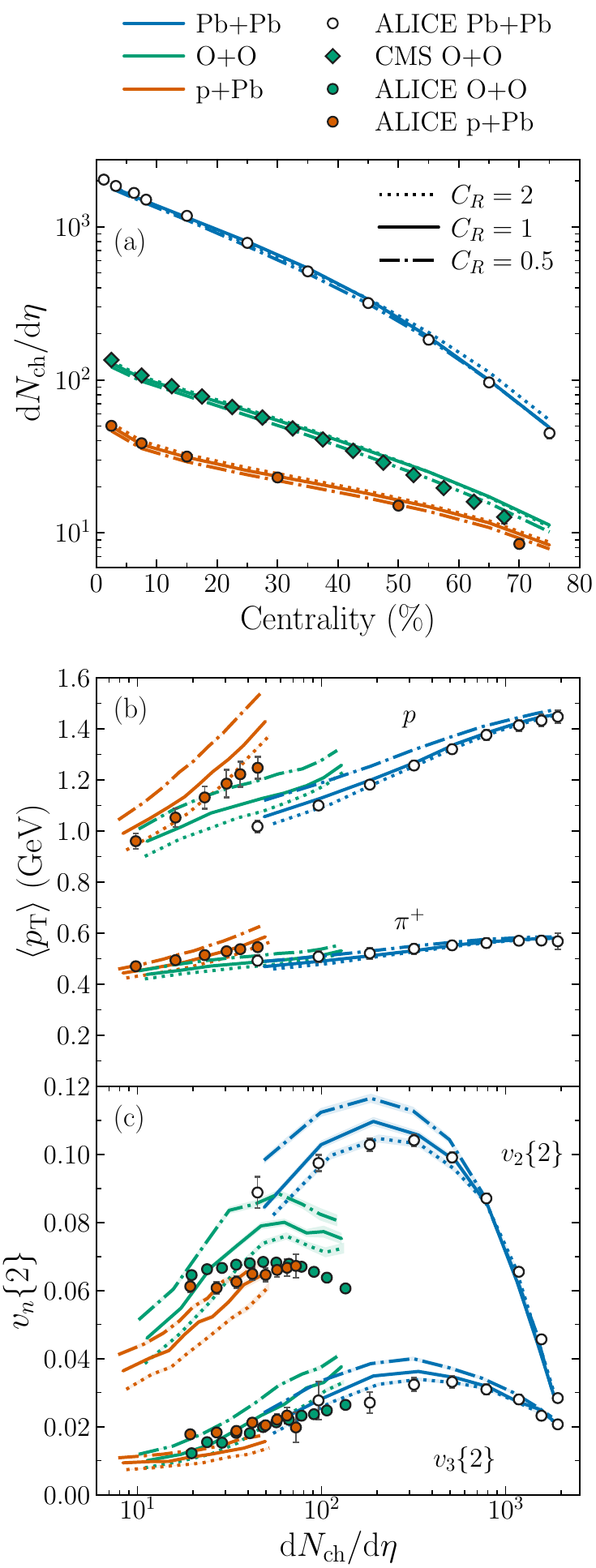}
  \caption{Sensitivity of final-state observables to the chemical equilibration rate parameter $C_R$ in $\mathrm{Pb}+\mathrm{Pb}$, $\mathrm{O}+\mathrm{O}$, and $\mathrm{p}+\mathrm{Pb}$ collisions at LHC energies: (a) the charged hadron multiplicity at mid-rapidity $\mathrm{d}N_{\mathrm{ch}}/\mathrm{d}\eta$ versus centrality; (b) The mean transverse momenta $\langle p_{\mathrm{T}}\rangle$ of $\pi^+$ and proton versus charged hadron multiplicity $\mathrm{d}N_{\mathrm{ch}}/\mathrm{d}\eta$; (c) charged hadron anisotropic flow coefficients $v_2\{2\}$ and $v_3\{2\}$ versus $\mathrm{d}N_{\mathrm{ch}}/\mathrm{d}\eta$. The experimental data are taken from Refs.~\cite{ALICE:2015juo, ALICE:2016ccg, ALICE:2019hno, ALICE:2013wgn, CMS:2026sai, ALICE:2025luc}.}
  \label{fig:finalstate_rate_scan}
\end{figure}

Figure~\ref{fig:finalstate_rate_scan} shows particle production and flow observables for three chemical equilibration rates controlled by the model parameter $C_R$ in $\mathrm{Pb}+\mathrm{Pb}$, $\mathrm{O}+\mathrm{O}$, and $p+\mathrm{Pb}$ collision systems.
We calibrate the model parameters (with the default choice of $C_R = 1$) to ALICE measurements in $\mathrm{Pb}+\mathrm{Pb}$ collisions at $\sqrt{s_{\mathrm{NN}}} = 5.02\;\mathrm{TeV}$. 
Model predictions for flow observables in $\mathrm{O}+\mathrm{O}$ and $\mathrm{p}+\mathrm{Pb}$ collisions are compared with recent measurements from the ALICE and CMS Collaborations.
We find that a larger chemical reaction rate $C_R$ leads to more viscous entropy/particle production, especially for small systems such as peripheral $\mathrm{Pb}+\mathrm{Pb}$, $\mathrm{O}+\mathrm{O}$, and $\mathrm{p}+\mathrm{Pb}$ collisions. Increasing $C_R$ from $0.5$ to $2$ raises $(\mathrm{d}N_{\mathrm{ch}}/\mathrm{d}\eta)$ by approximately $4 \%$ in central and $12 \%$ in peripheral $\mathrm{Pb}+\mathrm{Pb}$, $7–11\%$ in $\mathrm{O}+\mathrm{O}$, and $10–12\%$ in $\mathrm{p}+\mathrm{Pb}$ collisions. As alluded to in Fig.~\ref{fig:bulk_chem}, the system develops more hydrodynamic flow with a smaller (slower) chemical reaction rate coefficient $C_R$, which results in larger mean $p_{\rm T}$ and anisotropic flow coefficients for final-state hadrons shown in Figs.~\ref{fig:finalstate_rate_scan}b and ~\ref{fig:finalstate_rate_scan}c. 
Because the charged hadron multiplicity is proportional to the lifetime of the hydrodynamic phase, the sensitivity of the flow observables to $C_R$ increases with decreasing charged hadron multiplicity, while it shows less dependence on the collision system type. 
Therefore, the flow observables in small systems, such as $\mathrm{O}+\mathrm{O}$ and $\mathrm{p}+\mathrm{Pb}$ collisions, can provide phenomenological constraints on the chemical equilibration rate during the early pre-equilibrium phase.

\begin{figure}[t!]
  \centering
  \includegraphics[width=0.95\linewidth]{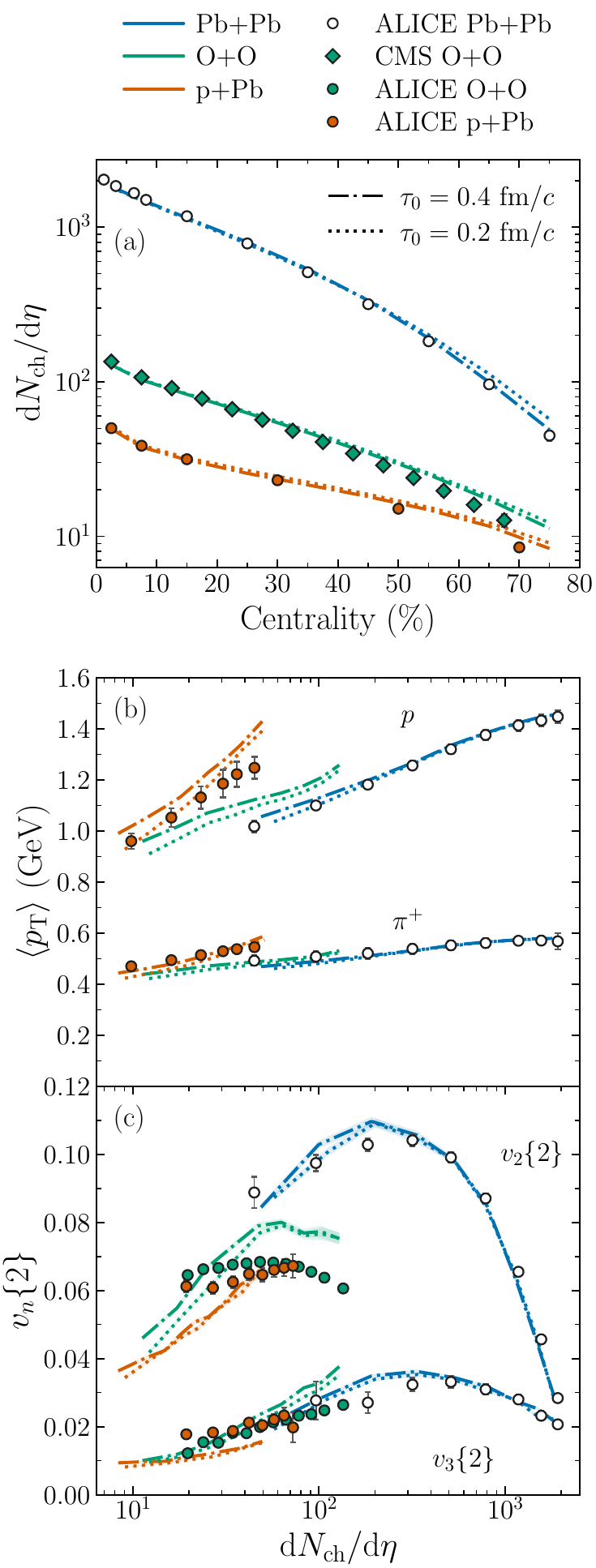}
  \caption{Similar to Fig.~\ref{fig:finalstate_rate_scan} but presenting the sensitivity of final-state observables to the hydrodynamic initialization time $\tau_0$ with chemical equilibration rate coefficient $C_R = 1$.}
  \label{fig:finalstate_tau0_scan}
\end{figure}

The dynamics of chemical equilibration during the early stage of the evolution also depends on the hydrodynamic starting time $\tau_0$ in our model. 
Figure~\ref{fig:finalstate_tau0_scan} investigates the sensitivity of final-state observables to the starting time of hydrodynamic evolution. 
A smaller $\tau_0$ lets the collision system produce viscous entropy earlier and yields more hadrons in the final state. 
This effect is sizable in peripheral $\mathrm{Pb}+\mathrm{Pb}$ collisions as shown in Fig.~\ref{fig:finalstate_tau0_scan}a. 
In contrast, simulations with a smaller $\tau_0$ evolve the system faster toward chemical equilibrium, which has lower pressure than the pure-gluon case. 
Therefore, it leads to less flow development during the early stage of hydrodynamics and results in smaller mean $p_{\rm T}$ and anisotropic flow coefficients for final-state hadrons, as shown in Figs.~\ref{fig:finalstate_tau0_scan}b and \ref{fig:finalstate_tau0_scan}c. 
The observables' sensitivity to the hydrodynamic starting time $\tau_0$ increases in collisions with small particle multiplicity (short hydrodynamic lifetime). 

\begin{figure}[t!]
  \centering
  \includegraphics[width=0.95\linewidth]{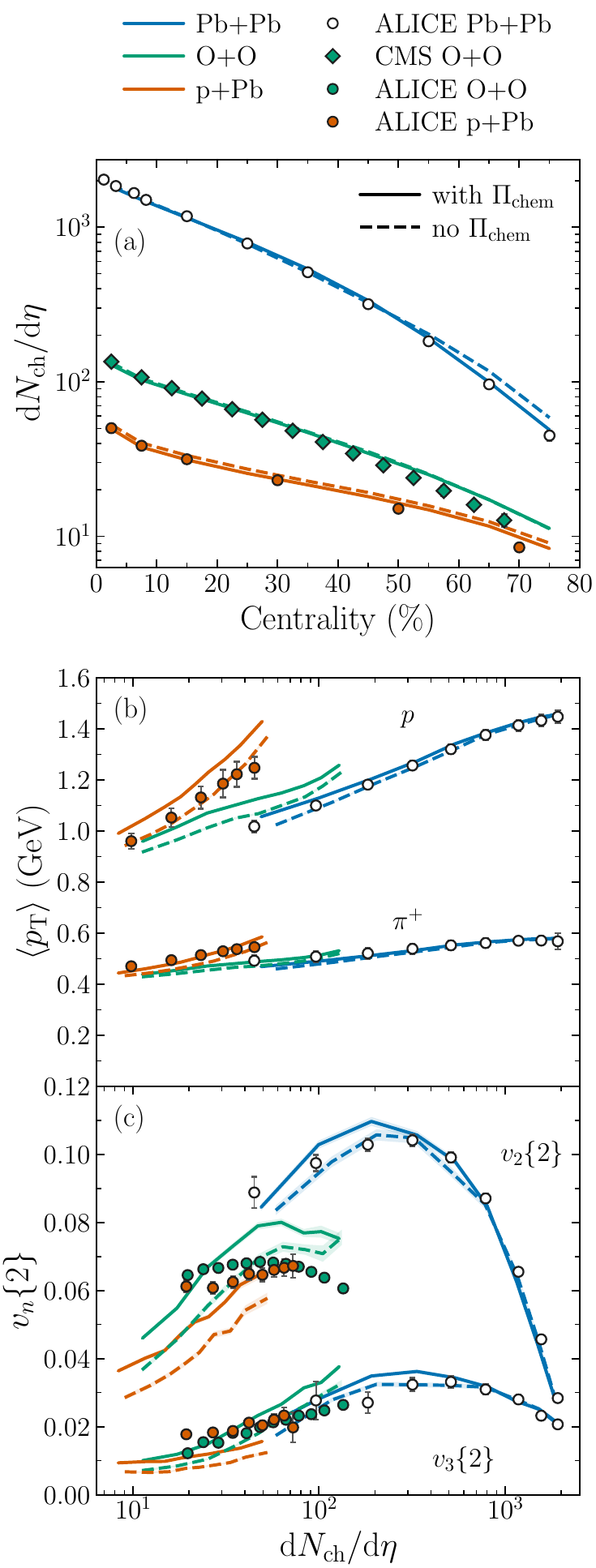}
  \caption{Similar to Fig.~\ref{fig:finalstate_rate_scan} but for a comparison with the standard setup without the pre-equilibrium effective bulk viscous pressure $\Pi_\mathrm{chem}$.}
  \label{fig:finalstate_with_without_pichem}
\end{figure}

Lastly, we compare simulation results in Fig.~\ref{fig:finalstate_with_without_pichem} with and without the pre-equilibrium effective bulk viscous pressure $\Pi_\mathrm{chem}$ (referred to as the standard setup~\cite{Mantysaari:2025tcg} below). 
This result quantifies the approximation of early-stage chemical equilibration with conventional bulk viscosity. 
We find negligible differences in particle multiplicity in central $\mathrm{Pb}+\mathrm{Pb}$ and $\mathrm{O}+\mathrm{O}$ collisions, while the hydrodynamic evolution with the effective $\Pi_\mathrm{chem}$ generates less viscous entropy production in peripheral $\mathrm{Pb}+\mathrm{Pb}$ and $\mathrm{p}+\mathrm{Pb}$ collisions.
Our model with $\Pi_\mathrm{chem}$ yields $5-10\%$ larger proton mean $p_{\rm T}$ and charged-hadron anisotropic flow coefficients than the standard setup, as shown in Fig.~\ref{fig:finalstate_with_without_pichem}b and~\ref{fig:finalstate_with_without_pichem}c. 
These results indicate that the bulk viscous pressure evolves faster in the standard setup compared to the case with $C_R = 1$.

Overall, we find that chemical equilibration effects during the early stage of evolution are small for long-lived collision systems, such as those in central $\mathrm{Pb}+\mathrm{Pb}$ collisions. 
Meanwhile, they produce $10\%$ modifications to the particles' mean $p_{\rm T}$ and anisotropic flow coefficients in small systems with $\d N_\mathrm{ch}/\d\eta < 100$ at the LHC. 

\section{Conclusion}
\label{Sec:Conclusion}

This study establishes a theoretical framework that describes microscopic chemical reaction processes for inelastic gluon fusion ($g + g \leftrightarrow q + \bar{q}$) using an effective macroscopic Israel–Stewart formulation of relativistic dissipative hydrodynamics. 
The chemical equilibration of quarks in the early, gluon-dominated stages of a heavy-ion collision acts as an effective, transient bulk viscous pressure driven by the non-conformality of the lattice QCD equation of state. 
The related effective transport coefficients -- the bulk relaxation time $\tau_{\Pi_\mathrm{chem}}$ and the specific bulk viscosity $\zeta_{\Pi_\mathrm{chem}}T/(e + P)$ -- are derived based on the microscopic chemical reaction rate, parametrized as $R = C_R T$.

Numerical simulations with this transient bulk viscous pressure show a hierarchy of system-size sensitivity in flow observables. 
In $\mathrm{Pb}+\mathrm{Pb}$ collisions at $\sqrt{s_{\mathrm{NN}}} = 5.02\;\mathrm{TeV}$, the chemical bulk pressure decays well before the dominant flow-building phase, leaving final-state observables unchanged at the sub-percent level. 
In intermediate $\mathrm{O}+\mathrm{O}$ collisions, a slow quark production rate $C_R = 0.5$ produces measurable effects: $\langle p_{\rm T} \rangle$ is enhanced by $5\%$ and $v_2\{2\}$ by $10\%$. 
In $\mathrm{p}+\mathrm{Pb}$ collisions, where the hydrodynamic lifetime is comparable to the chemical equilibration time, the effects are largest: $\langle p_{\rm T} \rangle$ increased by $8\%$ and $v_2\{2\}$ by $15\%$ with $C_R = 0.5$. 
A system-size scan of collective flow observables therefore provides a controlled experimental method to constrain the early-time chemical equilibration rate $C_R$. 
The recent $\mathrm{O}+\mathrm{O}$ collision programs at RHIC and LHC are particularly well suited for this purpose, as the intermediate system size provides optimal sensitivity to the chemical bulk pressure while maintaining the large particle multiplicities needed for precise flow measurements. 

Future extensions will explore the impacts of chemical non-equilibrium on electromagnetic observables (thermal photons and dileptons), extend to multi-scale chemical equilibration processes for strangeness, and perform systematic Bayesian inference on chemical equilibration rates with a system-size scan of collective flow observables.

The simulation results in this paper are produced by the \texttt{iEBE-MUSIC} framework. 
The specific version of the numerical framework and related parameter files are available in~\cite{akbarov_2026_23107614}.

\begin{acknowledgements}
This work is supported in part by the U.S. Department of Energy, Office of Science, Office of Nuclear Physics, under DOE Award No. DE-SC0021969.
C.S. acknowledges a DOE Office of Science Early Career Award. 
H.R. is supported by the Research Council of Finland, the Centre of Excellence in Quark Matter, and by the European Research Council (ERC, grant agreements No. ERC-2023-101123801 GlueSatLight and No. ERC-2018-ADG-835105 YoctoLHC), and by the European Union’s Marie Sk{\l}odowska-Curie Actions under Grant Agreement No. 101271866 (GSQUAD).
This research was done using computational resources provided by the Open Science Grid (OSG)~\cite{Pordes:2007zzb, Sfiligoi:2009cct, OSPool, OSDF}, which is supported by the National Science Foundation awards \#2030508 and \#2323298.
The content of this article does not reflect the official opinion of the European Union and responsibility for the information and views expressed therein lies entirely with the authors.
\end{acknowledgements}

\bibliography{references, non-inspires}

@misc{OSPool,
  doi = {10.21231/906P-4D78},
  url = {https://osg-htc.org/services/open_science_pool.html},
  author = {{OSG}},
  title = {OSPool},
  publisher = {OSG},
  year = {2006}
}

@misc{OSDF,
  doi = {10.21231/0KVZ-VE57},
  url = {https://osdf.osg-htc.org/},
  author = {{OSG}},
  title = {Open Science Data Federation},
  publisher = {OSG},
  year = {2015}
}

@software{akbarov_2026_23107614,
  author       = {Akbarov, Anar and
                  Roch, Hendrik and
                  Shen, Chun},
  title        = {iEBE-MUSIC simulation container and input
                   parameter files for chemical equilibration in
                   Pb+Pb, O+O and p+Pb collisions at LHC energies
                  },
  month        = oct,
  year         = 2026,
  publisher    = {Zenodo},
  version      = {2.0},
  doi          = {10.5281/zenodo.23107614},
  url          = {https://doi.org/10.5281/zenodo.23107614},
}

@article{Song:2010mg,
    author = "Song, Huichao and Bass, Steffen A. and Heinz, Ulrich and Hirano, Tetsufumi and Shen, Chun",
    title = "{200 A GeV Au+Au collisions serve a nearly perfect quark-gluon liquid}",
    eprint = "1011.2783",
    archivePrefix = "arXiv",
    primaryClass = "nucl-th",
    doi = "10.1103/PhysRevLett.106.192301",
    journal = "Phys. Rev. Lett.",
    volume = "106",
    pages = "192301",
    year = "2011",
    note = "[Erratum: Phys.Rev.Lett. 109, 139904 (2012)]"
}

@article{JETSCAPE:2026hdw,
    author = "Mankolli, A. and others",
    collaboration = "JETSCAPE",
    title = "{Longitudinal dynamics of large and small systems from a 3D Bayesian calibration of RHIC top-energy collision data}",
    eprint = "2601.17234",
    archivePrefix = "arXiv",
    primaryClass = "nucl-ex",
    doi = "10.1103/7kjw-vfgp",
    journal = "Phys. Rev. C",
    volume = "114",
    number = "1",
    pages = "014905",
    year = "2026"
}

@article{Jahan:2024wpj,
    author = "Jahan, Syed Afrid and Roch, Hendrik and Shen, Chun",
    title = "{Bayesian analysis of (3+1)D relativistic nuclear dynamics with the RHIC beam energy scan data}",
    eprint = "2408.00537",
    archivePrefix = "arXiv",
    primaryClass = "nucl-th",
    doi = "10.1103/PhysRevC.110.054905",
    journal = "Phys. Rev. C",
    volume = "110",
    number = "5",
    pages = "054905",
    year = "2024"
}

@article{Jahan:2025cbp,
    author = "Jahan, Syed Afrid and Roch, Hendrik and Shen, Chun",
    title = "{Bayesian model selection and uncertainty propagation for beam energy scan heavy-ion collisions}",
    eprint = "2507.11394",
    archivePrefix = "arXiv",
    primaryClass = "nucl-th",
    doi = "10.1103/y962-1lyg",
    journal = "Phys. Rev. C",
    volume = "113",
    number = "2",
    pages = "024919",
    year = "2026"
}

@article{Heinz:2013th,
    author = "Heinz, Ulrich and Snellings, Raimond",
    title = "{Collective flow and viscosity in relativistic heavy-ion collisions}",
    eprint = "1301.2826",
    archivePrefix = "arXiv",
    primaryClass = "nucl-th",
    doi = "10.1146/annurev-nucl-102212-170540",
    journal = "Ann. Rev. Nucl. Part. Sci.",
    volume = "63",
    pages = "123--151",
    year = "2013"
}

@article{Bernhard:2019bmu,
    author = "Bernhard, Jonah E. and Moreland, J. Scott and Bass, Steffen A.",
    title = "{Bayesian estimation of the specific shear and bulk viscosity of quark{\textendash}gluon plasma}",
    doi = "10.1038/s41567-019-0611-8",
    journal = "Nature Phys.",
    volume = "15",
    number = "11",
    pages = "1113--1117",
    year = "2019"
}

@article{ALICE:2022wpn,
    author = "Acharya, Shreyasi and others",
    collaboration = "ALICE",
    title = "{The ALICE experiment: a journey through QCD}",
    eprint = "2211.04384",
    archivePrefix = "arXiv",
    primaryClass = "nucl-ex",
    reportNumber = "CERN-EP-2022-227",
    doi = "10.1140/epjc/s10052-024-12935-y",
    journal = "Eur. Phys. J. C",
    volume = "84",
    number = "8",
    pages = "813",
    year = "2024"
}

@article{Gale:2013da,
    author = "Gale, Charles and Jeon, Sangyong and Schenke, Bjoern",
    title = "{Hydrodynamic Modeling of Heavy-Ion Collisions}",
    eprint = "1301.5893",
    archivePrefix = "arXiv",
    primaryClass = "nucl-th",
    doi = "10.1142/S0217751X13400113",
    journal = "Int. J. Mod. Phys. A",
    volume = "28",
    pages = "1340011",
    year = "2013"
}

@article{Busza:2018rrf,
    author = "Busza, Wit and Rajagopal, Krishna and van der Schee, Wilke",
    title = "{Heavy Ion Collisions: The Big Picture, and the Big Questions}",
    eprint = "1802.04801",
    archivePrefix = "arXiv",
    primaryClass = "hep-ph",
    reportNumber = "MIT-CTP-4892, MIT-CTP/4892",
    doi = "10.1146/annurev-nucl-101917-020852",
    journal = "Ann. Rev. Nucl. Part. Sci.",
    volume = "68",
    pages = "339--376",
    year = "2018"
}

@article{Berges:2020fwq,
    author = {Berges, J{\"u}rgen and Heller, Michal P. and Mazeliauskas, Aleksas and Venugopalan, Raju},
    title = "{QCD thermalization: Ab initio approaches and interdisciplinary connections}",
    eprint = "2005.12299",
    archivePrefix = "arXiv",
    primaryClass = "hep-th",
    reportNumber = "CERN-TH-2020-080",
    doi = "10.1103/RevModPhys.93.035003",
    journal = "Rev. Mod. Phys.",
    volume = "93",
    number = "3",
    pages = "035003",
    year = "2021"
}

@article{Shen:2020mgh,
    author = "Shen, Chun and Yan, Li",
    title = "{Recent development of hydrodynamic modeling in heavy-ion collisions}",
    eprint = "2010.12377",
    archivePrefix = "arXiv",
    primaryClass = "nucl-th",
    doi = "10.1007/s41365-020-00829-z",
    journal = "Nucl. Sci. Tech.",
    volume = "31",
    number = "12",
    pages = "122",
    month = "10",
    year = "2020"
}

@article{Schenke:2021mxx,
    author = {Schenke, Bj{\"o}rn},
    title = "{The smallest fluid on Earth}",
    eprint = "2102.11189",
    archivePrefix = "arXiv",
    primaryClass = "nucl-th",
    doi = "10.1088/1361-6633/ac14c9",
    journal = "Rept. Prog. Phys.",
    volume = "84",
    number = "8",
    pages = "082301",
    year = "2021"
}

@article{Noronha:2024dtq,
    author = {Noronha, Jorge and Schenke, Bj{\"o}rn and Shen, Chun and Zhao, Wenbin},
    title = "{Progress and challenges in small systems}",
    eprint = "2401.09208",
    archivePrefix = "arXiv",
    primaryClass = "nucl-th",
    doi = "10.1142/9789811294679_0004",
    journal = "Int. J. Mod. Phys. E",
    volume = "33",
    number = "06",
    pages = "2430005",
    year = "2024"
}

@article{Gelis:2009wh,
    author = "Gelis, F. and Lappi, T. and McLerran, L.",
    title = "{Glittering Glasmas}",
    eprint = "0905.3234",
    archivePrefix = "arXiv",
    primaryClass = "hep-ph",
    doi = "10.1016/j.nuclphysa.2009.07.004",
    journal = "Nucl. Phys. A",
    volume = "828",
    pages = "149--160",
    year = "2009"
}

@article{Schenke:2012wb,
    author = "Schenke, Bjoern and Tribedy, Prithwish and Venugopalan, Raju",
    title = "{Fluctuating Glasma initial conditions and flow in heavy ion collisions}",
    eprint = "1202.6646",
    archivePrefix = "arXiv",
    primaryClass = "nucl-th",
    doi = "10.1103/PhysRevLett.108.252301",
    journal = "Phys. Rev. Lett.",
    volume = "108",
    pages = "252301",
    year = "2012"
}

@article{Chesler:2010bi,
    author = "Chesler, Paul M. and Yaffe, Laurence G.",
    title = "{Holography and colliding gravitational shock waves in asymptotically AdS$_{5}$ spacetime}",
    eprint = "1011.3562",
    archivePrefix = "arXiv",
    primaryClass = "hep-th",
    doi = "10.1103/PhysRevLett.106.021601",
    journal = "Phys. Rev. Lett.",
    volume = "106",
    pages = "021601",
    year = "2011"
}

@article{Heller:2012km,
    author = "Heller, Michal P. and Mateos, David and van der Schee, Wilke and Trancanelli, Diego",
    title = "{Strong Coupling Isotropization of Non-Abelian Plasmas Simplified}",
    eprint = "1202.0981",
    archivePrefix = "arXiv",
    primaryClass = "hep-th",
    reportNumber = "ICCUB-12-009",
    doi = "10.1103/PhysRevLett.108.191601",
    journal = "Phys. Rev. Lett.",
    volume = "108",
    pages = "191601",
    year = "2012"
}

@article{Kurkela:2015qoa,
    author = "Kurkela, Aleksi and Zhu, Yan",
    title = "{Isotropization and hydrodynamization in weakly coupled heavy-ion collisions}",
    eprint = "1506.06647",
    archivePrefix = "arXiv",
    primaryClass = "hep-ph",
    reportNumber = "CERN-PH-TH-2015-142",
    doi = "10.1103/PhysRevLett.115.182301",
    journal = "Phys. Rev. Lett.",
    volume = "115",
    number = "18",
    pages = "182301",
    year = "2015"
}

@article{Kurkela:2018wud,
    author = {Kurkela, Aleksi and Mazeliauskas, Aleksas and Paquet, Jean-Fran{\c{c}}ois and Schlichting, S{\"o}ren and Teaney, Derek},
    title = "{Matching the Nonequilibrium Initial Stage of Heavy Ion Collisions to Hydrodynamics with QCD Kinetic Theory}",
    eprint = "1805.01604",
    archivePrefix = "arXiv",
    primaryClass = "hep-ph",
    doi = "10.1103/PhysRevLett.122.122302",
    journal = "Phys. Rev. Lett.",
    volume = "122",
    number = "12",
    pages = "122302",
    year = "2019"
}

@article{Kurkela:2018xxd,
    author = "Kurkela, Aleksi and Mazeliauskas, Aleksas",
    title = "{Chemical Equilibration in Hadronic Collisions}",
    eprint = "1811.03040",
    archivePrefix = "arXiv",
    primaryClass = "hep-ph",
    reportNumber = "CERN-TH-2018-238",
    doi = "10.1103/PhysRevLett.122.142301",
    journal = "Phys. Rev. Lett.",
    volume = "122",
    pages = "142301",
    year = "2019"
}

@article{Keegan:2015avk,
    author = "Keegan, Liam and Kurkela, Aleksi and Romatschke, Paul and van der Schee, Wilke and Zhu, Yan",
    title = "{Weak and strong coupling equilibration in nonabelian gauge theories}",
    eprint = "1512.05347",
    archivePrefix = "arXiv",
    primaryClass = "hep-th",
    reportNumber = "CERN-PH-TH-2015-304, MIT-CTP-4750",
    doi = "10.1007/JHEP04(2016)031",
    journal = "JHEP",
    volume = "04",
    pages = "031",
    year = "2016"
}

@article{Baier:2000sb,
    author = "Baier, R. and Mueller, Alfred H. and Schiff, D. and Son, D. T.",
    title = "{'Bottom up' thermalization in heavy ion collisions}",
    eprint = "hep-ph/0009237",
    archivePrefix = "arXiv",
    doi = "10.1016/S0370-2693(01)00191-5",
    journal = "Phys. Lett. B",
    volume = "502",
    pages = "51--58",
    year = "2001"
}

@article{Liu:2015nwa,
    author = "Liu, Jia and Shen, Chun and Heinz, Ulrich",
    title = "{Pre-equilibrium evolution effects on heavy-ion collision observables}",
    eprint = "1504.02160",
    archivePrefix = "arXiv",
    primaryClass = "nucl-th",
    doi = "10.1103/PhysRevC.91.064906",
    journal = "Phys. Rev. C",
    volume = "91",
    number = "6",
    pages = "064906",
    year = "2015",
    note = "[Erratum: Phys.Rev.C 92, 049904 (2015)]"
}

@article{Kurkela:2018vqr,
    author = {Kurkela, Aleksi and Mazeliauskas, Aleksas and Paquet, Jean-Fran{\c{c}}ois and Schlichting, S{\"o}ren and Teaney, Derek},
    title = "{Effective kinetic description of event-by-event pre-equilibrium dynamics in high-energy heavy-ion collisions}",
    eprint = "1805.00961",
    archivePrefix = "arXiv",
    primaryClass = "hep-ph",
    doi = "10.1103/PhysRevC.99.034910",
    journal = "Phys. Rev. C",
    volume = "99",
    number = "3",
    pages = "034910",
    year = "2019"
}

@article{Schenke:2020mbo,
    author = "Schenke, Bjoern and Shen, Chun and Tribedy, Prithwish",
    title = "{Running the gamut of high energy nuclear collisions}",
    eprint = "2005.14682",
    archivePrefix = "arXiv",
    primaryClass = "nucl-th",
    doi = "10.1103/PhysRevC.102.044905",
    journal = "Phys. Rev. C",
    volume = "102",
    number = "4",
    pages = "044905",
    year = "2020"
}

@article{Borsanyi:2013bia,
    author = "Borsanyi, Szabocls and Fodor, Zoltan and Hoelbling, Christian and Katz, Sandor D. and Krieg, Stefan and Szabo, Kalman K.",
    title = "{Full result for the QCD equation of state with 2+1 flavors}",
    eprint = "1309.5258",
    archivePrefix = "arXiv",
    primaryClass = "hep-lat",
    doi = "10.1016/j.physletb.2014.01.007",
    journal = "Phys. Lett. B",
    volume = "730",
    pages = "99--104",
    year = "2014"
}

@article{HotQCD:2014kol,
    author = "Bazavov, A. and others",
    collaboration = "HotQCD",
    title = "{Equation of state in ( 2+1 )-flavor QCD}",
    eprint = "1407.6387",
    archivePrefix = "arXiv",
    primaryClass = "hep-lat",
    reportNumber = "BNL-105928-2014-JA",
    doi = "10.1103/PhysRevD.90.094503",
    journal = "Phys. Rev. D",
    volume = "90",
    pages = "094503",
    year = "2014"
}

@article{daSilva:2022xwu,
    author = "da Silva, Tiago Nunes and Chinellato, David D. and Giannini, Andr{\'e} V. and Ferreira, Maur{\'\i}cio N. and Denicol, Gabriel S. and Hippert, Maur{\'\i}cio and Luzum, Matthew and Noronha, Jorge and Takahashi, Jun",
    collaboration = "ExTrEMe",
    title = "{Prehydrodynamic evolution in large and small systems}",
    eprint = "2211.10561",
    archivePrefix = "arXiv",
    primaryClass = "nucl-th",
    doi = "10.1103/PhysRevC.107.044901",
    journal = "Phys. Rev. C",
    volume = "107",
    number = "4",
    pages = "044901",
    year = "2023"
}

@article{Schenke:2019pmk,
    author = "Schenke, Bjoern and Shen, Chun and Tribedy, Prithwish",
    title = "{Hybrid Color Glass Condensate and hydrodynamic description of the Relativistic Heavy Ion Collider small system scan}",
    eprint = "1908.06212",
    archivePrefix = "arXiv",
    primaryClass = "nucl-th",
    doi = "10.1016/j.physletb.2020.135322",
    journal = "Phys. Lett. B",
    volume = "803",
    pages = "135322",
    year = "2020"
}

@article{JETSCAPE:2020mzn,
    author = "Everett, D. and others",
    collaboration = "JETSCAPE",
    title = "{Multisystem Bayesian constraints on the transport coefficients of QCD matter}",
    eprint = "2011.01430",
    archivePrefix = "arXiv",
    primaryClass = "hep-ph",
    doi = "10.1103/PhysRevC.103.054904",
    journal = "Phys. Rev. C",
    volume = "103",
    number = "5",
    pages = "054904",
    year = "2021"
}

@article{Nijs:2020roc,
    author = {Nijs, Govert and van der Schee, Wilke and G{\"u}rsoy, Umut and Snellings, Raimond},
    title = "{Bayesian analysis of heavy ion collisions with the heavy ion computational framework Trajectum}",
    eprint = "2010.15134",
    archivePrefix = "arXiv",
    primaryClass = "nucl-th",
    reportNumber = "CERN-TH-2020-175, MIT-CTP/5251",
    doi = "10.1103/PhysRevC.103.054909",
    journal = "Phys. Rev. C",
    volume = "103",
    number = "5",
    pages = "054909",
    year = "2021"
}

@article{Borghini:2024kll,
    author = "Borghini, Nicolas and Krupczak, Renata and Roch, Hendrik",
    title = "{Comparing matching prescriptions between pre-equilibrium and hydrodynamic models in high-energy nuclear collisions}",
    eprint = "2407.10634",
    archivePrefix = "arXiv",
    primaryClass = "nucl-th",
    doi = "10.1140/epjc/s10052-024-13509-8",
    journal = "Eur. Phys. J. C",
    volume = "84",
    number = "10",
    pages = "1128",
    year = "2024"
}

@article{Chiu:2021muk,
    author = "Chiu, Cheng and Shen, Chun",
    title = "{Exploring theoretical uncertainties in the hydrodynamic description of relativistic heavy-ion collisions}",
    eprint = "2103.09848",
    archivePrefix = "arXiv",
    primaryClass = "nucl-th",
    doi = "10.1103/PhysRevC.103.064901",
    journal = "Phys. Rev. C",
    volume = "103",
    number = "6",
    pages = "064901",
    year = "2021"
}

@article{Chiu:2025mau,
    author = "Chiu, Cheng and Denicol, Gabriel and Luzum, Matthew and Shen, Chun",
    title = "{A Resummed Hydrodynamic Description of Relativistic Heavy-ion Collisions}",
    eprint = "2508.05292",
    archivePrefix = "arXiv",
    primaryClass = "nucl-th",
    doi = "10.1088/1361-6471/aea878",
    month = "8",
    year = "2025",
    journal = "",
}

@article{Gordeev:2025vog,
    author = "Gordeev, Andrew and Bass, Steffen A. and Mueller, Berndt and Paquet, Jean-Francois",
    title = "{Quark flavor equilibration of the quark-gluon plasma}",
    eprint = "2501.06433",
    archivePrefix = "arXiv",
    primaryClass = "hep-ph",
    doi = "10.1103/w7tz-mlh7",
    journal = "Phys. Rev. C",
    volume = "113",
    number = "1",
    pages = "014906",
    year = "2026"
}

@article{Vovchenko:2016ijt,
    author = {Vovchenko, V. and Karpenko, Iu. A. and Gorenstein, M. I. and Satarov, L. M. and Mishustin, I. N. and K{\"a}mpfer, B. and Stoecker, H.},
    title = "{Electromagnetic probes of a pure-glue initial state in nucleus-nucleus collisions at energies available at the CERN Large Hadron Collider}",
    eprint = "1604.06346",
    archivePrefix = "arXiv",
    primaryClass = "nucl-th",
    doi = "10.1103/PhysRevC.94.024906",
    journal = "Phys. Rev. C",
    volume = "94",
    number = "2",
    pages = "024906",
    year = "2016"
}

@article{Gavassino:2023xkt,
    author = "Gavassino, Lorenzo and Noronha, Jorge",
    title = "{Relativistic bulk rheology: From neutron star mergers to viscous cosmology}",
    eprint = "2305.04119",
    archivePrefix = "arXiv",
    primaryClass = "gr-qc",
    doi = "10.1103/PhysRevD.109.096040",
    journal = "Phys. Rev. D",
    volume = "109",
    number = "9",
    pages = "096040",
    year = "2024"
}

@article{Yang:2023ogo,
    author = "Yang, Yumu and Hippert, Mauricio and Speranza, Enrico and Noronha, Jorge",
    title = "{Far-from-equilibrium bulk-viscous transport coefficients in neutron star mergers}",
    eprint = "2309.01864",
    archivePrefix = "arXiv",
    primaryClass = "nucl-th",
    doi = "10.1103/PhysRevC.109.015805",
    journal = "Phys. Rev. C",
    volume = "109",
    number = "1",
    pages = "015805",
    year = "2024"
}

@article{Israel:1979wp,
    author = "Israel, W. and Stewart, J. M.",
    title = "{Transient relativistic thermodynamics and kinetic theory}",
    doi = "10.1016/0003-4916(79)90130-1",
    journal = "Annals Phys.",
    volume = "118",
    pages = "341--372",
    year = "1979"
}

@article{Moreland:2015dvc,
    author = "Moreland, J. Scott and Soltz, Ron A.",
    title = "{Hydrodynamic simulations of relativistic heavy-ion collisions with different lattice quantum chromodynamics calculations of the equation of state}",
    eprint = "1512.02189",
    archivePrefix = "arXiv",
    primaryClass = "nucl-th",
    doi = "10.1103/PhysRevC.93.044913",
    journal = "Phys. Rev. C",
    volume = "93",
    number = "4",
    pages = "044913",
    year = "2016"
}

@article{Borsanyi:2012ve,
    author = "Borsanyi, Sz. and Endrodi, G. and Fodor, Z. and Katz, S. D. and Szabo, K. K.",
    title = "{Precision SU(3) lattice thermodynamics for a large temperature range}",
    eprint = "1204.6184",
    archivePrefix = "arXiv",
    primaryClass = "hep-lat",
    reportNumber = "WUB-12-09",
    doi = "10.1007/JHEP07(2012)056",
    journal = "JHEP",
    volume = "07",
    pages = "056",
    year = "2012"
}

@article{Mantysaari:2025tcg,
    author = {M{\"a}ntysaari, Heikki and Schenke, Bj{\"o}rn and Shen, Chun and Zhao, Wenbin},
    title = "{Collision-Energy Dependence in Heavy-Ion Collisions from Nonlinear QCD Evolution}",
    eprint = "2502.05138",
    archivePrefix = "arXiv",
    primaryClass = "nucl-th",
    doi = "10.1103/gf4y-p5j7",
    journal = "Phys. Rev. Lett.",
    volume = "135",
    number = "2",
    pages = "022302",
    year = "2025"
}

@article{Kowalski:2003hm,
    author = "Kowalski, Henri and Teaney, Derek",
    title = "{An Impact parameter dipole saturation model}",
    eprint = "hep-ph/0304189",
    archivePrefix = "arXiv",
    doi = "10.1103/PhysRevD.68.114005",
    journal = "Phys. Rev. D",
    volume = "68",
    pages = "114005",
    year = "2003"
}

@article{Schlichting:2014ipa,
    author = {Schlichting, S{\"o}ren and Schenke, Bj{\"o}rn},
    title = "{The shape of the proton at high energies}",
    eprint = "1407.8458",
    archivePrefix = "arXiv",
    primaryClass = "hep-ph",
    doi = "10.1016/j.physletb.2014.10.068",
    journal = "Phys. Lett. B",
    volume = "739",
    pages = "313--319",
    year = "2014"
}

@article{Mantysaari:2022sux,
    author = {M{\"a}ntysaari, Heikki and Salazar, Farid and Schenke, Bj{\"o}rn},
    title = "{Nuclear geometry at high energy from exclusive vector meson production}",
    eprint = "2207.03712",
    archivePrefix = "arXiv",
    primaryClass = "hep-ph",
    doi = "10.1103/PhysRevD.106.074019",
    journal = "Phys. Rev. D",
    volume = "106",
    number = "7",
    pages = "074019",
    year = "2022"
}

@article{Paquet:2015lta,
    author = {Paquet, Jean-Fran{\c{c}}ois and Shen, Chun and Denicol, Gabriel S. and Luzum, Matthew and Schenke, Bj{\"o}rn and Jeon, Sangyong and Gale, Charles},
    title = "{Production of photons in relativistic heavy-ion collisions}",
    eprint = "1509.06738",
    archivePrefix = "arXiv",
    primaryClass = "hep-ph",
    doi = "10.1103/PhysRevC.93.044906",
    journal = "Phys. Rev. C",
    volume = "93",
    number = "4",
    pages = "044906",
    year = "2016"
}

@article{Denicol:2012cn,
    author = "Denicol, G. S. and Niemi, H. and Molnar, E. and Rischke, D. H.",
    title = "{Derivation of transient relativistic fluid dynamics from the Boltzmann equation}",
    eprint = "1202.4551",
    archivePrefix = "arXiv",
    primaryClass = "nucl-th",
    doi = "10.1103/PhysRevD.85.114047",
    journal = "Phys. Rev. D",
    volume = "85",
    pages = "114047",
    year = "2012",
    note = "[Erratum: Phys.Rev.D 91, 039902 (2015)]"
}

@article{Denicol:2014vaa,
    author = "Denicol, G. S. and Jeon, S. and Gale, C.",
    title = "{Transport Coefficients of Bulk Viscous Pressure in the 14-moment approximation}",
    eprint = "1403.0962",
    archivePrefix = "arXiv",
    primaryClass = "nucl-th",
    doi = "10.1103/PhysRevC.90.024912",
    journal = "Phys. Rev. C",
    volume = "90",
    number = "2",
    pages = "024912",
    year = "2014"
}

@article{Huovinen:2012is,
    author = "Huovinen, Pasi and Petersen, Hannah",
    title = "{Particlization in hybrid models}",
    eprint = "1206.3371",
    archivePrefix = "arXiv",
    primaryClass = "nucl-th",
    doi = "10.1140/epja/i2012-12171-9",
    journal = "Eur. Phys. J. A",
    volume = "48",
    pages = "171",
    year = "2012"
}

@article{Shen:2014vra,
    author = "Shen, Chun and Qiu, Zhi and Song, Huichao and Bernhard, Jonah and Bass, Steffen and Heinz, Ulrich",
    title = "{The iEBE-VISHNU code package for relativistic heavy-ion collisions}",
    eprint = "1409.8164",
    archivePrefix = "arXiv",
    primaryClass = "nucl-th",
    doi = "10.1016/j.cpc.2015.08.039",
    journal = "Comput. Phys. Commun.",
    volume = "199",
    pages = "61--85",
    year = "2016"
}

@article{Bass:1998ca,
    author = "Bass, S. A. and others",
    title = "{Microscopic models for ultrarelativistic heavy ion collisions}",
    eprint = "nucl-th/9803035",
    archivePrefix = "arXiv",
    doi = "10.1016/S0146-6410(98)00058-1",
    journal = "Prog. Part. Nucl. Phys.",
    volume = "41",
    pages = "255--369",
    year = "1998"
}

@article{Bleicher:1999xi,
    author = "Bleicher, M. and others",
    title = "{Relativistic hadron hadron collisions in the ultrarelativistic quantum molecular dynamics model}",
    eprint = "hep-ph/9909407",
    archivePrefix = "arXiv",
    doi = "10.1088/0954-3899/25/9/308",
    journal = "J. Phys. G",
    volume = "25",
    pages = "1859--1896",
    year = "1999"
}

@article{ALICE:2015juo,
    author = "Adam, Jaroslav and others",
    collaboration = "ALICE",
    title = "{Centrality Dependence of the Charged-Particle Multiplicity Density at Midrapidity in Pb-Pb Collisions at $\sqrt{s_{\rm NN}}$ = 5.02 TeV}",
    eprint = "1512.06104",
    archivePrefix = "arXiv",
    primaryClass = "nucl-ex",
    reportNumber = "CERN-PH-EP-2015-324, ALICE-PUBLIC-2015-008",
    doi = "10.1103/PhysRevLett.116.222302",
    journal = "Phys. Rev. Lett.",
    volume = "116",
    number = "22",
    pages = "222302",
    year = "2016"
}

@article{ALICE:2016ccg,
    author = "Adam, Jaroslav and others",
    collaboration = "ALICE",
    title = "{Anisotropic flow of charged particles in Pb-Pb collisions at $\sqrt{s_{\rm NN}}=5.02$ TeV}",
    eprint = "1602.01119",
    archivePrefix = "arXiv",
    primaryClass = "nucl-ex",
    reportNumber = "CERN-EP-2016-018",
    doi = "10.1103/PhysRevLett.116.132302",
    journal = "Phys. Rev. Lett.",
    volume = "116",
    number = "13",
    pages = "132302",
    year = "2016"
}

@article{ALICE:2013wgn,
    author = "Abelev, Betty Bezverkhny and others",
    collaboration = "ALICE",
    title = "{Multiplicity Dependence of Pion, Kaon, Proton and Lambda Production in p-Pb Collisions at $\sqrt{s_{NN}}$ = 5.02 TeV}",
    eprint = "1307.6796",
    archivePrefix = "arXiv",
    primaryClass = "nucl-ex",
    reportNumber = "CERN-PH-EP-2013-135",
    doi = "10.1016/j.physletb.2013.11.020",
    journal = "Phys. Lett. B",
    volume = "728",
    pages = "25--38",
    year = "2014"
}

@article{Mantysaari:2025cos,
    author = {M{\"a}ntysaari, Heikki and Schenke, Bj{\"o}rn and Shen, Chun and Zhao, Wenbin},
    title = "{Perturbative high-energy evolution in the IP-Glasma initial state}",
    eprint = "2508.20432",
    archivePrefix = "arXiv",
    primaryClass = "nucl-th",
    doi = "10.1051/epjconf/202636409006",
    journal = "EPJ Web Conf.",
    volume = "364",
    pages = "09006",
    year = "2026"
}

@article{Mantysaari:2025kls,
    author = {M{\"a}ntysaari, Heikki and Schenke, Bj{\"o}rn and Shen, Chun and Zhao, Wenbin},
    title = "{Impact of QCD energy evolution on observables in heavy-ion collisions}",
    eprint = "2511.03588",
    archivePrefix = "arXiv",
    primaryClass = "nucl-th",
    doi = "10.1103/qsd7-zwmw",
    journal = "Phys. Rev. C",
    volume = "113",
    number = "3",
    pages = "034914",
    year = "2026"
}

@article{Du:2025hyk,
    author = {Du, Xiaojian and Schlichting, S{\"o}ren and Zhu, Jie},
    title = "{(3+1)D event-by-event preequilibrium dynamics in heavy-ion collisions}",
    eprint = "2512.08007",
    archivePrefix = "arXiv",
    primaryClass = "hep-ph",
    doi = "10.1103/hpzj-xx3h",
    journal = "Phys. Rev. D",
    volume = "113",
    number = "9",
    pages = "096009",
    year = "2026"
}

@article{CMS:2026sai,
    author = "Belyaev, Andrey and others",
    collaboration = "CMS",
    title = "{Centrality dependence of charged-hadron pseudorapidity distributions in oxygen-oxygen collisions at $\sqrt{s_\mathrm{NN}}$ = 5.36 TeV}",
    eprint = "2606.02285",
    archivePrefix = "arXiv",
    primaryClass = "nucl-ex",
    reportNumber = "CMS-HIN-25-010, CERN-EP-2026-153",
    month = "6",
    year = "2026",
    journal = "",
}

@article{ALICE:2019hno,
    author = "Acharya, Shreyasi and others",
    collaboration = "ALICE",
    title = "{Production of charged pions, kaons, and (anti-)protons in Pb-Pb and inelastic $pp$ collisions at $\sqrt {s_{NN}}$ = 5.02 TeV}",
    eprint = "1910.07678",
    archivePrefix = "arXiv",
    primaryClass = "nucl-ex",
    reportNumber = "CERN-EP-2019-208",
    doi = "10.1103/PhysRevC.101.044907",
    journal = "Phys. Rev. C",
    volume = "101",
    number = "4",
    pages = "044907",
    year = "2020"
}

@article{ALICE:2025luc,
    author = "Abualrob, Ibrahim Jaser and others",
    collaboration = "ALICE",
    title = "{Evidence of Nuclear Geometry-Driven Anisotropic Flow in O+O and Ne+Ne Collisions at sNN=5.36{\,}{\,}TeV}",
    eprint = "2509.06428",
    archivePrefix = "arXiv",
    primaryClass = "nucl-ex",
    reportNumber = "CERN-EP-2025-203",
    doi = "10.1103/gymp-vp87",
    journal = "Phys. Rev. Lett.",
    volume = "137",
    number = "8",
    pages = "082301",
    year = "2026"
}

@article{Pordes:2007zzb,
    author = "Pordes, Ruth and others",
    editor = "Keyes, David E.",
    title = "{The Open Science Grid}",
    reportNumber = "FERMILAB-CONF-07-217-CD",
    doi = "10.1088/1742-6596/78/1/012057",
    journal = "J. Phys. Conf. Ser.",
    volume = "78",
    pages = "012057",
    year = "2007"
}

@article{Sfiligoi:2009cct,
    author = "Sfiligoi, Igor and Bradley, Daniel C. and Holzman, Burt and Mhashilkar, Parag and Padhi, Sanjay and Wurthwrin, Frank",
    title = "{The pilot way to Grid resources using glideinWMS}",
    reportNumber = "FERMILAB-CONF-09-373-CD",
    doi = "10.1109/CSIE.2009.950",
    journal = "WRI World Congress",
    volume = "2",
    pages = "428--432",
    year = "2009"
}

\end{document}